\documentclass[aps,prl,showkeys,preprint,preprintnumbers,superscriptaddress]{revtex4-1}
\def\^{\hat}
\def\~{\tilde}
\def\3h{{3\over 2}}

\def\v#1{{\bf#1}}

\def\E{{\v E}}
\def\ee{{\v e}}
\def\h{{\v h}}

\def\H{{\v H}}

\def\B{{\v B}}

\def\ep{\epsilon}

\def\pd#1#2{{{\partial #1}\over{\partial #2}}}

\def\div{\nabla\cdot}
\def\gradt{\nabla_{t}}
\def\divt{\nabla_{t}\cdot}
\def\curl{\nabla\times}

\def\curlt{\nabla_{t}\times}

\def\eqn#1$${\eqno{{\rm #1}}$$}

\def\ep{\epsilon}

\def\~{\tilde}

\def\^{\hat}

\def\d#1{\overline{\overline{#1}}}

\def\XXint#1#2#3{{\setbox0=\hbox{$#1{#2#3}{\int}$}
     \vcenter{\hbox{$#2#3$}}\kern-.5\wd0}}

\usepackage{amssymb}
\usepackage{amsthm}
\usepackage{graphics}
\usepackage{graphicx}
\usepackage{amsfonts}
\usepackage[cmex10]{amsmath}
\usepackage{array}
\usepackage{natbib}
\usepackage{lineno}
\usepackage{epstopdf}
\usepackage{subfigure}
\usepackage{epstopdf}
\allowdisplaybreaks
\begin{document}

%\preprint{Physical Review Letters}
\title{A Necessary and Sufficient Condition for Having Independent TE and TM Modes in an Anisotropic Waveguide}


\author{W. Jiang}
\email[]{jwmathphy@163.com}
%\affiliation{}

\affiliation{Institute of Electromagnetics and Acoustics, and Department of Electronic Science, Xiamen University, Xiamen 361005, China}
\author{J. Liu}
\email[]{liujie190484@stu.xmu.edu.cn}
\affiliation{Institute of Electromagnetics and Acoustics, and Department of Electronic Science, Xiamen University, Xiamen 361005, China}
\author{Q. Li}
\affiliation{School of Science, Beijing Technology and Business University, Beijing 100048, China}
\author{T. Xia}
\affiliation{School of Mathematics and Statistics, Guizhou University of Finance and Economics, Guiyang 550025, China}
\author{Y. H. Xu}
\affiliation{ School of Finance, Nanjing Audit University, Jiangsu 211815, China}
\author{N. Liu}
\affiliation{Institute of Electromagnetics and Acoustics, and Department of Electronic Science, Xiamen University, Xiamen 361005, China}
\author{Q. H. Liu}
\email[]{qhliu@duke.edu}
\affiliation{Department of Electrical and Computer Engineering,
Duke University, Durham, NC 27708 USA}

\date{\today}
\begin{abstract}
Whether there exist independent transverse electric (TE) and transverse magnetic (TM) modes in a metallic waveguide filled with an anisotropic medium is a fundamental question in electromagnetics waveguide theory, but so far no definitive answers have been published. This paper establishes a necessary and sufficient condition for having independent TE and TM modes in a waveguide filled with a homogeneous lossless anisotropic medium based on both waveguide theory in electromagnetics and basic knowledge in mathematics. Moreover, for the independent TE modes, we prove the propagation constants obtained from both the longitudinal scalar magnetic field stimulation and the transverse vector electric field stimulation are the same; for the independent TM modes, the propagation constants obtained from both the longitudinal scalar electric field stimulation and the transverse vector magnetic field stimulation are the same. This necessary and sufficient condition is a new theoretical result in electromagnetic waveguide theory, and is expected to be valuable for the design of waveguides filled with anisotropic media.
\end{abstract}

%\pacs{}
%\pacs{#1}
\maketitle
\section{1. Introduction}

Electromagnetic waveguide is one of the most important source-free devices in physics and engineering. It guides electromagnetic waves
along the longitudinal ($z$) direction with little loss. For example, waveguide antennas can conceive multicolor coherent
manipulation schemes of individual emitters \cite{Mermillod}.

According to the waveguide theory in electromagnetics \cite{Collin1960,Balanis1989}, it is well-known that there exist independent TE (transverse electric) modes and TM (transverse magnetic) modes in a metallic (perfect electric conductor) waveguide filled with a  homogeneous lossless isotropic medium. Moreover, for the independent TE modes, we know that these independent TE modes can be stimulated by employing longitudinal scalar magnetic field $h_{z}$, which has to solve an eigenvalue problem about Laplace operator with a Neumann boundary condition, and can also be stimulated by employing transverse vector electric field $\ee_{t}$, which needs to solve an eigenvalue problem about curl-curl operator with a Dirichlet boundary condition. Similarly, for the independent TM modes, we can use longitudinal scalar electric field $e_{z}$ to stimulate them, which has to solve an eigenvalue problem about the Laplace operator with a Dirichlet boundary condition, and can also use transverse vector magnetic field $\h_{t}$ to stimulate them, which needs to solve an eigenvalue problem about the curl-curl operator with two Neumann boundary conditions.

If the metallic waveguide is filled with a homogeneous lossless anisotropic medium, under what conditions it will have independent TE modes and TM modes? This question is both interesting and fundamental for electromagnetic wave field theory, but it has not been answered in literature. In general, if the medium in the waveguide is fully anisotropic, independent TE modes and TM modes cannot exist in this waveguide because they are coupled together. How about the anisotropic medium along the $xy$ plane only? The answer is that only under some special conditions, there exist independent TE modes and TM modes. This paper finds and proves a necessary and sufficient condition for having independent TE modes and TM modes in an anisotropic waveguide. To prove this necessary and sufficient condition, we need to use electromagnetic theory and basic knowledge in mathematics, for example, linear algebra theory \cite{Strang2006}, mathematical analysis \cite{Rudin1964}, some identities in field theory \cite{adamson2007introduction}, simple complex number theory \cite{Agarwal2011}, etc.

The outline of this paper is as follows. In Sec. 2, we review the governing equations for waveguide problem according to the waveguide theory. In Sec. 3, we establish a necessary condition for having independent TE modes and TM modes in the waveguide based on both waveguide theory in electromagnetics and basic mathematics. In Sec. 4, we prove that the necessary condition established by us is also a sufficient condition. Finally, based on the conclusion in Sec. 3 and Sec. 4, we propose the necessary and sufficient condition for having independent TE modes and TM modes in the waveguide filled with a homogenous lossless anisotropic medium.

\section{2. Governing Equations for Waveguide Problem}

 Let $\Gamma$ be the cross section of the metallic (PEC) waveguide; let $\^n$ be the outward normal unit vector on the PEC boundary $\partial\Gamma$ of the cross section $\Gamma$. As usual, the cross section $\Gamma$ is bounded. Because of the complexity of the cross section $\Gamma$, the boundary $\partial\Gamma$ may be not connected, i.e., there are multiple PEC walls (for example, inner and outer conductors of a coaxial cable). In this case, there exist TEM modes. In practical applications, waveguide walls are often made of PEC, with the notable exception of dielectric waveguides that are unbounded.

 The waveguide problem is a so-called 2.5-dimensional problem, because the medium in the waveguide is two-dimensional (thus invariant in the $z$ direction), while the electromagnetic field in the waveguide is
three-dimensional. This paper only treats the waveguide problem filled with a homogenous lossless anisotropic medium.
As a waveguide filled with a fully anisotropic medium does not support independent TE and TM modes, we suppose that the permittivity and permeability tensors are of the following form:
\begin{eqnarray}
  \d{\ep}&=&\left[
  \begin{array}{cc}
    \d{\ep}_{t} & 0 \\
    0 & \ep_{zz} \\
  \end{array}
\right],\quad\quad
\d{\ep}_{t}=\left[\begin{array}{cc}
    a_{1} &b_{1}+c_{1}j  \\
    b_{1}-c_{1}j  & d_{1} \\
  \end{array}
\right],
\label{ep}\\
\d{\mu}&=&\left[
  \begin{array}{cc}
    \d{\mu}_{t} & 0 \\
    0 & \mu_{zz} \\
  \end{array}
\right],\quad\quad
\d{\mu}_{t}=\left[\begin{array}{cc}
    a_{2} &b_{2}+c_{2}j  \\
    b_{2}-c_{2}j  & d_{2} \\
  \end{array}
\right],\label{mu}
\end{eqnarray}
where both $\d{\ep}$ and $\d{\mu}$ are constant matrices. Because the medium considered here is lossless, we assume that $\d{\ep}$ and $\d{\mu}$ are two positive definite Hermitian matrices \cite{Chew1990}, i.e.,
$$\d{\ep}_{t}^{\dagger}=\d{\ep}_{t}, \quad\d{\mu}_{t}^{\dagger}=\d{\mu}_{t}, ~~\ep_{zz}>0, ~~\mu_{zz}>0,$$
where both $\d{\ep}_{t}$ and $\d{\mu}_{t}$ are also two positive definite Hermitian matrices, and the superscript $\dagger$ denotes complex conjugate transpose of a matrix. According to the properties of a positive definite Hermitian matrix, we assume that $a_{i}>0,~d_{i}>0~(i=1,2)$.

In a usual waveguide problem, the operating frequency $f>0$ is given, then $\omega=2\pi f>0$. We need to solve for the propagation wavenumber $k_z$ ($k_{z}>0$) of several propagation physical modes:
\begin{equation}\label{guide1}
    \E(x,y,z)=\ee(x,y)e^{-jk_{z}z},~\H(x,y,z)=\h(x,y)e^{-jk_{z}z},
\end{equation}
where $\ee(x,y)$ and $\h(x,y)$ are two three-dimensional vectors only dependent on the transversal coordinates $(x,y)$, while independent with longitudinal coordinate $z$. Note that
\begin{eqnarray}
    &~&\ee(x,y)=\ee_{t}+\^ze_{z},\quad\h(x,y)=\h_{t}+\^zh_{z},\label{guide2} \\ &~&\nabla=\nabla_{t}+\^z\pd{}{z},\quad\nabla_{t}=\^x\pd{}{x}+\^y\pd{}{y}.\label{guidddde2}
\end{eqnarray}
For the waveguide problem, we need to solve the source-free Maxwell's equations:
\begin{eqnarray*}
  \curl\E&=&-j\omega\d{\mu}\H, \\
  \curl\H&=& j\omega\d{\ep}\E,\\
  \div(\d{\ep}\E)&=&0,\\
  \div(\d{\mu}\H)&=&0,
\end{eqnarray*}
where we have assumed that the time-harmonic factor is $e^{j\omega t}$. Substituting (\ref{ep}-\ref{guidddde2}) into the above Maxwell's equations, we obtain
the following partial differential equations (PDEs):
\begin{eqnarray}
  \curlt\ee_{t} &=& -j\omega\mu_{zz}h_z\^z\label{eqn31} \\
  -\^z\times\nabla_{t}e_z-jk_z\^z\times\ee_{t}&=& -j\omega\d{\mu}_{t}\h_{t} \label{eqn32}\\
  \curlt\h_{t} &=& j\omega\ep_{zz}e_z\^z\label{eqn33} \\
  -\^z\times\nabla_{t}h_z-jk_z\^z\times\h_{t}&=& j\omega\d{\ep}_{t}\ee_{t} \label{eqn34}\\
  \divt(\d{\ep}_{t}\ee_{t})&=&jk_{z}\ep_{zz}e_z\label{addeqn35}\\
  \divt(\d{\mu}_{t}\h_{t})&=&jk_{z}\mu_{zz}h_z.\label{addeqn36}
\end{eqnarray}
Set $A=\left[
       \begin{array}{cc}
         0 & -1 \\
         1 & 0 \\
       \end{array}
     \right]$, then $A^2=-\d{I}_{2\times2}$,
where $\d{I}_{2\times2}$ is identity matrix. We change $\^z\times$ in equations (\ref{eqn32}) and (\ref{eqn34}) into $A$, then we get
\begin{eqnarray}
\nabla_{t}e_{z}+jk_{z}\ee_{t}&=&-j\omega A\d{\mu}_{t}\h_{t},\label{eqn35}\\
\nabla_{t}h_{z}+jk_{z}\h_{t}&=&j\omega A\d{\ep}_{t}\ee_{t}.\label{eqn36}
\end{eqnarray}
From equations (\ref{eqn35}) and (\ref{eqn36}), we can obtain
\begin{eqnarray}
(-\omega^2A\d{\mu}_{t}A\d{\ep}_{t}-k_{z}^2\d{I}_{2\times2})\ee_{t}
=-jk_{z}\nabla_{t}e_{z}+j\omega A\d{\mu}_{t}\nabla_{t}h_{z}, \label{eqn37}\\
(-\omega^2A\d{\ep}_{t}A\d{\mu}_{t}-k_{z}^2\d{I}_{2\times2})\h_{t}
=-jk_{z}\nabla_{t}h_{z}-j\omega A\d{\ep}_{t}\nabla_{t}e_{z}. \label{eqn38}
\end{eqnarray}
Set $N_{1}=-\omega^2A\d{\mu}_{t}A\d{\ep}_{t}-k_{z}^2\d{I}_{2\times2}$ and $N_{2}=-\omega^2A\d{\ep}_{t}A\d{\mu}_{t}-k_{z}^2\d{I}_{2\times2}$. If $\det(N_{1})\neq0$ and $\det(N_{2})\neq0$, then one has
 \begin{eqnarray}
\ee_{t}
=-jk_{z}N_{1}^{-1}\nabla_{t}e_{z}+j\omega N_{1}^{-1}A\d{\mu}_{t}\nabla_{t}h_{z}, \label{eeqn37}\\
\h_{t}
=-jk_{z}N_{2}^{-1}\nabla_{t}h_{z}-j\omega N_{2}^{-1} A\d{\ep}_{t}\nabla_{t}e_{z}. \label{eeqn38}
\end{eqnarray}
Substituting (\ref{eeqn37}) and (\ref{eeqn38}) into equations (\ref{eqn31}) and (\ref{eqn33}) respectively yields
\begin{eqnarray}
k_{z}\curlt(N_{1}^{-1}\nabla_{t}e_{z})-\omega \curlt(N_{1}^{-1}A\d{\mu}_{t}\nabla_{t}h_{z})&=&\omega\mu_{zz}h_{z}\^z \label{main1}\\
k_{z}\curlt(N_{2}^{-1}\nabla_{t}h_{z})+\omega \curlt(N_{2}^{-1}A\d{\ep}_{t}\nabla_{t}e_{z})&=&-\omega\ep_{zz}e_{z}\^z \label{main2}
\end{eqnarray}

In this paper, we treat the waveguide problem with the PEC outer boundary condition. Thus, from the electromagnetic theory \cite{Balanis1989}, we have $$\^n\times\E=0\mbox{~on~}\partial\Gamma,\quad\^n\cdot\B=0\mbox{~on~}\partial\Gamma.$$

\section{3. Necessary Condition}

It is well-known that the waveguide has independent TE modes if and only if $e_{z}=0$ and $h_{z}\neq0$, and it has independent TM modes if and only if $h_{z}=0$ and $e_{z}\neq0$.

For the independent TE modes, we take $e_{z}=0$ in equations (\ref{main1})
and (\ref{main2}) respectively, then the necessary condition is
\begin{eqnarray}
-\curlt(N_{1}^{-1}A\d{\mu}_{t}\nabla_{t}h_{z})&=&\mu_{zz}h_{z}\^z, \label{main3}\\
\curlt(N_{2}^{-1}\nabla_{t}h_{z})&=&0. \label{main4}
\end{eqnarray}
\indent For the independent TM modes, we take $h_{z}=0$ in equations (\ref{main1})
and (\ref{main2}) respectively, then the necessary condition is
\begin{eqnarray}
-\curlt(N_{2}^{-1}A\d{\ep}_{t}\nabla_{t}e_{z})&=&\ep_{zz}e_{z}\^z,\label{main5}\\
 \curlt(N_{1}^{-1}\nabla_{t}e_{z})&=&0.\label{main6}
\end{eqnarray}
Set $C^{2}(\Gamma)=\{\varphi: \varphi \mbox{~is twice continuous derivative~} \mbox{in}~\Gamma\}$.\\
{\bf{Lemma 1:}} Suppose that $X_{2\times 2}$ is a nonzero constant matrix,
$\curlt(X\nabla_{t}\psi)=0~\mbox{in~}{\Gamma}$ for any scalar field $\psi\in{C^{2}(\Gamma)}$ if and only if $X=\Lambda\d{I}_{2\times2}$, where $\Lambda$ is a nonzero constant.\\
Proof: If $X=\Lambda\d{I}_{2\times2}$,  then $\curlt(X\nabla_{t}\psi)=\curlt(\Lambda\d{I}_{2\times2}\nabla_{t}\psi)
=\Lambda\curlt(\nabla_{t}\psi)=0$. Conversely, if $\curlt(X\nabla_{t}\psi)=0~\mbox{in~}{\Gamma}$ for any scalar field $\psi\in{C^{2}(\Gamma)}$, assume that
\begin{equation*}
    X=\left[
        \begin{array}{cc}
          x_1 & x_2 \\
          x_3 & x_4\\
        \end{array}
      \right],
\end{equation*}
then
$$x_3\pd{^2\psi}{x^2}+(x_4-x_1)\pd{^2\psi}{x\partial y}-x_2\pd{^2\psi}{y^2}=0,~\forall\psi\in{C^{2}(\Gamma)}.$$
Due to the arbitrariness in the choice of the function $\psi$, we can get $x_3=x_2=0$, and $x_1=x_4=\Lambda\neq0$, which shows that $X=\Lambda\d{I}_{2\times2}\neq0$. $\qed$ \\
{\bf{Theorem 1:}} The matrix $N_1$ is a scalar matrix if and only if the matrix $N_2$ is a scalar matrix, and in this case $N_1=N_2$.\\
\indent Proof: Because $N_{1}=-\omega^2A\d{\mu}_{t}A\d{\ep}_{t}-k_{z}^2\d{I}_{2\times2}$,  $N_{2}=-\omega^2A\d{\ep}_{t}A\d{\mu}_{t}-k_{z}^2\d{I}_{2\times2}$, and $k_{z}^2\d{I}_{2\times2}$ has already been a scalar matrix, we only need to prove that $A\d{\mu}_{t}A\d{\ep}_{t}$ is a scalar matrix if and only if $A\d{\ep}_{t}A\d{\mu}_{t}$ is a scalar matrix, and in this case $A\d{\ep}_{t}A\d{\mu}_{t}=A\d{\mu}_{t}A\d{\ep}_{t}$.
\begin{eqnarray*}
A\d{\mu}_{t}A\d{\ep}_{t}&=&\left[
  \begin{array}{cc}
    0 & -1 \\
    1 & 0 \\
  \end{array}
\right]\left[\begin{array}{cc}
    a_{2} &b_{2}+c_{2}j  \\
    b_{2}-c_{2}j  & d_{2} \\
  \end{array}
\right]\left[
  \begin{array}{cc}
    0 & -1 \\
    1 & 0 \\
  \end{array}
\right]\left[\begin{array}{cc}
    a_{1} &b_{1}+c_{1}j  \\
    b_{1}-c_{1}j  & d_{1} \\
  \end{array}
\right]\\
&=&\left[
     \begin{array}{cc}
       (b_1 - c_{1}j)(b_{2} - c_{2}j)- a_{1}d_{2} &~~ d_{1}(b_{2} - c_{2}j) - d_{2}(b_{1} + c_{1}j)\\
        a_{1}(b_{2} + c_{2}j) - a_{2}(b_{1} - c_{1}j) &~~ (b_{1} + c_{1}j)(b_{2} + c_{2}j) - a_{2}d_{1} \\
     \end{array}
   \right],\\
A\d{\ep}_{t}A\d{\mu}_{t}&=&\left[
  \begin{array}{cc}
    0 & -1 \\
    1 & 0 \\
  \end{array}
\right]\left[\begin{array}{cc}
    a_{1} &b_{1}+c_{1}j  \\
    b_{1}-c_{1}j  & d_{1} \\
  \end{array}
\right]\left[
  \begin{array}{cc}
    0 & -1 \\
    1 & 0 \\
  \end{array}
\right]\left[\begin{array}{cc}
    a_{2} &b_{2}+c_{2}j  \\
    b_{2}-c_{2}j  & d_{2} \\
  \end{array}
\right]\\
&=&\left[
     \begin{array}{cc}
       (b_{1} - c_{1}j)(b_{2} - c_{2}j) - a_{2}d_{1} & ~~d_{2}(b_{1} - c_{1}j) - d_{1}(b_{2} + c_{2}j)\\
        a_{2}(b_{1} + c_{1}j) - a_{1}(b_{2} - c_{2}j) &~~ (b_{1} + c_{1}j)(b_{2} + c_{2}j) - a_{1}d_{2} \\
     \end{array}
   \right].\\
\end{eqnarray*}
$A\d{\mu}_{t}A\d{\ep}_{t}$ is a scalar matrix if and only if it satisfies the following equations:
\begin{equation}\label{applendex1}
\left\{ \begin{aligned}
         (b_1 - c_{1}j)(b_{2} - c_{2}j)- a_{1}d_{2}&=(b_{1} + c_{1}j)(b_{2} + c_{2}j) - a_{2}d_{1}\\
         d_{1}(b_{2} - c_{2}j) - d_{2}(b_{1} + c_{1}j)&=0\\
         a_{1}(b_{2} + c_{2}j) - a_{2}(b_{1} - c_{1}j)&=0.
        \end{aligned} \right.
\end{equation}
$A\d{\ep}_{t}A\d{\mu}_{t}$ is a scalar matrix if and only if it satisfies the following equations:
\begin{equation}\label{applendex2}
\left\{ \begin{aligned}
         (b_{1} - c_{1}j)(b_{2} - c_{2}j) - a_{2}d_{1}&=(b_{1} + c_{1}j)(b_{2} + c_{2}j) - a_{1}d_{2}\\
         d_{2}(b_{1} - c_{1}j) - d_{1}(b_{2} + c_{2}j)&=0\\
         a_{2}(b_{1} + c_{1}j) - a_{1}(b_{2} - c_{2}j)&=0.
        \end{aligned} \right.
\end{equation}
In fact, equations (\ref{applendex1}) are equivalent to equations (\ref{applendex2}), because equations (\ref{applendex1}) and (\ref{applendex2}) are mutually complex conjugate. Therefore
$A\d{\mu}_{t}A\d{\ep}_{t}$ is a scalar matrix if and only
if $A\d{\ep}_{t}A\d{\mu}_{t}$ is a scalar matrix, and in this case it is easy to prove that $A\d{\mu}_{t}A\d{\ep}_{t}=A\d{\ep}_{t}A\d{\mu}_{t}$. The proof of Theorem 1 is thus completed. $\qed$\\
\indent Hence, based on Lemma 1, if the waveguide has independent TE modes, we must have the conclusion that
$N_2$ in equation (\ref{main4}) is a scalar matrix. Similarly, if the waveguide has independent TM modes, we must have the conclusion that $N_1$ in equation (\ref{main6}) is a scalar matrix. According to Theorem 1, we know that $N_1=N_2$. Hence, $N_1$ and $N_2$ in equations (\ref{main3}) and (\ref{main4}) are the same; the scalar matrix $N_1$ and $N_2$ in the equations (\ref{main5}) and (\ref{main6}) are also the same.
Based on the proof process of Theorem 1, we can achieve the necessary condition for the independent TE modes and TM modes in the waveguide, and that is:

Condition $\mathfrak{C}$:

  \begin{eqnarray}\label{important1}
 \d{\ep}=\left[\begin{array}{ccc}
    \ep & (b+cj)\ep &0 \\
    (b-cj)\ep  & a^2\ep &0\\
    0  & 0 &\ep_{zz}\\
  \end{array}
\right],~~
\d{\mu}=\left[\begin{array}{ccc}
    \mu &(b-cj)\mu &0 \\
    (b+cj)\mu  & a^2\mu &0\\
    0  & 0 &\mu_{zz}\\
  \end{array}
\right],
\end{eqnarray}
where $\d{\ep}$ and $\d{\mu}$ are two positive definite Hermitian matrices.
From the positive definiteness of (\ref{important1}), we conclude that $\ep>0$, $\mu>0$ and $a^2-(b^2+c^2)>0$ according to linear algebra theory \cite{Strang2006}.

\section{4. Sufficient Condition}

In this section, we prove that the above necessary Condition $\mathfrak{C}$ is also a sufficient condition. Suppose that the medium in the waveguide has already satisfies Condition $\mathfrak{C}$, then we can define
\begin{equation}\label{define1}
 -\omega^2\^z\times\d{\ep}_{t}\^z\times\d{\mu}_{t}=
 -\omega^2\^z\times\d{\mu}_{t}\^z\times\d{\ep}_{t}=k^2\d{I}_{2\times2},
\end{equation}
 where $k^2=\omega^2\big[a^2-(b^2+c^2)\big]\ep\mu>0$. In fact, the above identity (\ref{define1}) can be seen as an operator equation, and $\^z\times$ can be seen as a rotational transformation in the $xy$ plane, which is a linear operator. Below we find the sufficient conditions for the existence of independent TE and TM modes.

\subsection{4.1 Independent TE Modes}

The governing equations for simulating independent TE modes by using longitudinal component $h_z$ reads:

Seek $k_{z}\in{\mathbb{R}}$, $h_z\neq 0$, such that
\begin{equation} \label{3eqs1}
\left\{ \begin{aligned}
         -\divt(\d{\mu}_{t}\nabla_{t}h_{z}) &= k_{t}^2\mu_{zz}h_z=(k^2-k_{z}^2)\mu_{zz}h_z~~\mbox{in}~\Gamma\\
                  \^n\cdot(\d{\mu}_{t}\gradt h_z)&=0~~\mbox{on}~\partial\Gamma,
                          \end{aligned} \right.
\end{equation}
where $k_{t}^{2}=k^{2}-k_{z}^{2}$. Once any eigen-pair $(k_{z},h_z)$ is solved in PDE (\ref{3eqs1}), then
\begin{eqnarray*}
\ee_{t}=\frac{j\omega}{k_{t}^2}\^z\times(\d{\mu}_{t}\nabla_{t}h_{z}), \quad\h_{t}=-\frac{jk_{z}}{k_{t}^2}\nabla_{t}h_{z}.
\end{eqnarray*}
To simulate independent TE modes by using transverse component $\ee_t$, the governing equations can be written as:

Find $k_{z}\in{\mathbb{R}}$, $\ee_t\neq 0$, such that
\begin{equation} \label{3eqs2}
\left\{ \begin{aligned}
         \curlt\big(\mu_{zz}^{-1}\curlt\ee_t\big)-\omega^2\d{\ep}_{t}\ee_{t}&
         =k_{z}^2\big(\^z\times\d{\mu}_{t}^{-1}\^z\times\ee_t\big)~~\mbox{in}~\Gamma, \\
         \divt(\d{\ep}_{t}\ee_t)&=0~~\mbox{in}~\Gamma,\\
         \^n\times\ee_t&=0~~\mbox{on}~\partial\Gamma.
        \end{aligned} \right.
\end{equation}
Once any eigen-pair $(k_{z},\ee_t)$ is solved in PDEs (\ref{3eqs2}), then
\begin{eqnarray*}
\h_{t}=\frac{k_{z}\d{\mu}_{t}^{-1}(\^z\times\ee_t)}{\omega},\quad h_z=\^z\cdot\frac{j\curlt\ee_t}{\omega\mu_{zz}}.
\end{eqnarray*}

For the equations (\ref{3eqs1}) and (\ref{3eqs2}), we achieve the following two important Lemmas.

\noindent{\bf{Lemma 2.}} If $(k_{z},h_{z})$ is an eigen-pair of PDE (\ref{3eqs1}),  then $(k_{z}, \frac{j\omega}{k_{t}^2}\^z\times(\d{\mu}_{t}\nabla_{t}h_{z}))$ is also an eigen-pair of PDEs (\ref{3eqs2}). \\
Proof:
Firstly we verify the correctness of the first equation in PDEs (\ref{3eqs2}),
\begin{eqnarray*}
  &~&\curlt\bigg(\mu_{zz}^{-1}\curlt\ee_t\bigg)-k_{z}^2(\^z\times\d{\mu}_{t}^{-1}\^z\times\ee_{t})-\omega^2\d{\ep}_{t}\ee_{t}\\
  &=&\frac{j\omega}{k_{t}^2}\bigg[\mu_{zz}^{-1}\curlt\bigg(\curlt\big(\^z\times(\d{\mu}_{t}\nabla_{t}h_{z})
  \big)\bigg)-k_{z}^2\big(\^z\times\d{\mu}_{t}^{-1}\^z\times\^z\times(\d{\mu}_{t}\nabla_{t}h_{z})\big)-\omega^2\d{\ep}_{t}\^z\times(\d{\mu}_{t}\nabla_{t}h_{z})\bigg]\\
   &=&\frac{j\omega}{k_{t}^2}\bigg[\mu_{zz}^{-1}\curlt\bigg(\curlt\big(\^z\times(\d{\mu}_{t}\nabla_{t}h_{z})
  \big)\bigg)+k_{z}^2\big(\^z\times\nabla_{t}h_{z}\big)-\omega^2\d{\ep}_{t}\^z\times(\d{\mu}_{t}\nabla_{t}h_{z})\bigg]\\
  &=&\frac{j\omega}{k_{t}^2}\bigg[\mu_{zz}^{-1}\curlt\bigg(\^z\cdot\big(\divt(\d{\mu}_{t}\nabla_{t}h_{z})\big)\bigg)+k_{z}^2\big(\^z\times\nabla_{t}h_{z}\big)-\^z\times(-\omega^2\^z\times\d{\ep}_{t}\^z\times\d{\mu}_{t})\nabla_{t}h_{z}\bigg]\\
  &=&\frac{j\omega}{k_{t}^2}\bigg[-\mu_{zz}^{-1}\^z\times\bigg(\gradt\big(\divt(\d{\mu}_{t}\nabla_{t}h_{z})\big)\bigg)+k_{z}^2\big(\^z\times\nabla_{t}h_{z}\big)-k^2\big(\^z\times\nabla_{t}h_{z}\big)\bigg]\\
  &=&-\frac{j\omega\mu_{zz}^{-1}}{k_{t}^2}\bigg[\^z\times\bigg(\gradt\big(\divt(\d{\mu}_{t}\nabla_{t}h_{z})\big)\bigg)-(k_{z}^2-k^2)\big(\^z\times\nabla_{t}(\mu_{zz}h_{z})\big)\bigg]\\
    &=&-\frac{j\omega\mu_{zz}^{-1}}{k_{t}^2}\bigg[\^z\times\gradt\bigg(\divt(\d{\mu}_{t}\nabla_{t}h_{z})+(k^2-k_{z}^2)\mu_{zz}h_{z}\bigg)\bigg]=0.\\
\end{eqnarray*}
Secondly we verify the correctness of the second equation in PDEs (\ref{3eqs2}),
\begin{eqnarray*}
&&\divt\big(\^z\times\d{\mu}_{t}^{-1}\^z\times\ee_t\big)=\divt\big(\^z\times\d{\mu}_{t}^{-1}\^z\times\frac{j\omega}{k_{t}^2}\^z\times(\d{\mu}_{t}\nabla_{t}h_{z})\big)\\
&=&-\frac{j\omega}{k_{t}^2}\divt\big(\^z\times\d{\mu}_{t}^{-1}(\d{\mu}_{t}\nabla_{t}h_{z})\big)=-\frac{j\omega}{k_{t}^2}\divt\big(\^z\times\nabla_{t}h_{z})\big)\\
&=&\frac{j\omega}{k_{t}^2}\bigg(\^z\cdot(\curlt\gradt h_z)\bigg)=0.\\
\end{eqnarray*}
From the first equation in PDEs (\ref{3eqs2}), we have
$$\ep_{t}\ee_{t}=\frac{1}{\omega^2}\bigg[\curlt\bigg(\mu_{zz}^{-1}\curlt\ee_t\bigg)-
k_{z}^2\bigg(\^z\times\d{\mu}_{t}^{-1}\^z\times\ee_t\bigg)\bigg].$$
Taking the divergence for the above equation, then we arrive at
\begin{eqnarray*}
\divt(\d{\ep}_{t}\ee_t)&=&\frac{1}{\omega^2}\divt\bigg[\curlt\bigg(\mu_{zz}^{-1}\curlt\ee_t\bigg)-
k_{z}^2\bigg(\^z\times\d{\mu}_{t}^{-1}\^z\times\ee_t\bigg)\bigg]\\
&=&\frac{1}{\omega^2}\divt\bigg[\curlt\bigg(\mu_{zz}^{-1}\curlt\ee_t\bigg)\bigg]
-\frac{k_{z}^2}{\omega^2}\divt\big(\^z\times\d{\mu}_{t}^{-1}\^z\times\ee_t\big)=0,
\end{eqnarray*}
which validates the second equation in PDEs (\ref{3eqs2}). \\
\indent Finally we validate the boundary condition in PDEs (\ref{3eqs2}).
\begin{eqnarray*}
  \^n\times\ee_t&=&\^n\times\big(\frac{j\omega}{k_{t}^2}\^z\times(\d{\mu}_{t}\nabla_{t}h_{z})\big)\\
   &=&\frac{j\omega}{k_{t}^2}\^n\times\big(\^z\times(\d{\mu}_{t}\nabla_{t}h_{z})\big) \\
   &=&\frac{j\omega}{k_{t}^2}\bigg(\^z\big(\^n\cdot(\d{\mu}_{t}\nabla_{t}h_{z})\big)-\d{\mu}_{t}\nabla_{t}h_{z}(\^n\cdot\^z)\bigg)\\
   &=&0,
\end{eqnarray*}
which validates boundary condition in PDEs (\ref{3eqs2}). The proof of Lemma 2 is thus completed. $\qed$\\
%\begin{widetext}
%\end{widetext}

\noindent{\bf{Lemma 3.}} If $(k_{z},\ee_{t})$ is an eigen-pair of PDEs (\ref{3eqs2}),  then $(k_{z}, \^z\cdot\frac{j\curlt\ee_t}{\omega\mu_{zz}})$ is also an eigen-pair of PDE (\ref{3eqs1}). \\
Proof: Firstly we verify the equation in PDE (\ref{3eqs1}),
\begin{eqnarray*}
   &~&\divt(\d{\mu}_{t}\gradt h_{z})+(k^2-k_{z}^2)\mu_{zz} h_{z}\\
   &=&\divt\big(\d{\mu}_{t}\gradt(\^z\cdot\frac{j\curlt\ee_{t}}{\omega\mu_{zz}})\big)
   +(k^2-k_{z}^2)\mu_{zz}\big(\^z\cdot\frac{j\curlt\ee_{t}}{\omega\mu_{zz}}\big)\\
   &=&\frac{j}{\omega}\bigg[\divt\bigg(\d{\mu}_{t}\gradt\big(\^z\cdot(\mu_{zz}^{-1}\curlt\ee_{t})\big)\bigg)
   +(k^2-k_{z}^2)\^z\cdot\curlt\ee_{t}\bigg]\\
   &=&\frac{j}{\omega}\bigg[\divt\bigg(\d{\mu}_{t}\^z\times\big(\curlt(\mu_{zz}^{-1}\curlt\ee_{t})\big)\bigg)
   +(k^2-k_{z}^2)\^z\cdot\curlt\ee_{t}\bigg]\\
   &=&\frac{j}{\omega}\bigg[\divt\bigg(\d{\mu}_{t}\^z\times\big(\omega^2\d{\ep}_{t}\ee_{t}+k_{z}^2(\^z\times\d{\mu}_{t}^{-1}\^z\times\ee_{t})\big)\bigg)
   +(k^2-k_{z}^2)\^z\cdot\curlt\ee_{t}\bigg]\\
   &=&\frac{j}{\omega}\bigg[\divt\big(\d{\mu}_{t}\^z\times(\omega^2\d{\ep}_{t}\ee_{t})\big)+k_{z}^2\divt(\d{\mu}_{t}\^z\times\^z\times\d{\mu}_{t}^{-1}\^z\times\ee_{t})
   +(k^2-k_{z}^2)\^z\cdot\curlt\ee_{t}\bigg]\\
   &=&\frac{j}{\omega}\bigg[\divt\big(\d{\mu}_{t}\^z\times(\omega^2\d{\ep}_{t}\ee_{t})\big)-k_{z}^2\divt(\^z\times\ee_{t})
   +(k^2-k_{z}^2)\^z\cdot\curlt\ee_{t}\bigg]\\
   &=&\frac{j}{\omega}\bigg[\divt\big(\omega^2\d{\mu}_{t}\^z\times(\d{\ep}_{t}\ee_{t})\big)
   +k^2\^z\cdot\curlt\ee_{t}\bigg]\\
   &=&\frac{j}{\omega}\bigg[\divt\big(-\^z\times\omega^2\^z\times\d{\mu}_{t}\^z\times(\d{\ep}_{t}\ee_{t})\big)
   +k^2\^z\cdot\curlt\ee_{t}\bigg]\\
   &=&\frac{j}{\omega}\bigg[k^2\divt\big(\^z\times\ee_{t}\big)
   +k^2\^z\cdot\curlt\ee_{t}\bigg]
   =\frac{jk^2}{\omega}\bigg[\divt\big(\^z\times\ee_{t}\big)-\divt\big(\^z\times\ee_{t}\big)\bigg]=0.
\end{eqnarray*}
Finally we verify the boundary condition in PDE (\ref{3eqs1}),
\begin{eqnarray*}
&~&\^n\cdot(\d{\mu}_{t}\gradt h_z)=
\frac{j}{\omega}\^n\cdot\big(\d{\mu}_{t}\gradt(\mu_{zz}^{-1}\curlt\ee_{t})\big)\\
&=&\frac{j}{\omega}\^n\cdot\bigg(\d{\mu}_{t}\^z\times\big(\curlt(\mu_{zz}^{-1}\curlt\ee_{t})\big)\bigg)\\
&=&\frac{j}{\omega}\^n\cdot\bigg(\d{\mu}_{t}\^z\times\big(\omega^2\d{\ep}_{t}\ee_{t}+
k_{z}^2(\^z\times\d{\mu}_{t}^{-1}\^z\times\ee_{t})\big)\bigg)\\
&=&\frac{j}{\omega}\^n\cdot\big(\d{\mu}_{t}\^z\times\omega^2\d{\ep}_{t}\ee_{t}-k_{z}^2(\^z\times\ee_{t})\big)\\
&=&\frac{j}{\omega}\^n\cdot\big(\^z\times(-\omega^2\^z\times\d{\mu}_{t}\^z\times\d{\ep}_{t})\ee_{t}-k_{z}^2(\^z\times\ee_{t})\big)\\
&=&\frac{j}{\omega}\^n\cdot\big(k^2(\^z\times\ee_{t})-k_{z}^2(\^z\times\ee_{t})\big)=\frac{jk_{t}^2}{\omega}\^n\cdot(\^z\times\ee_{t})\\
&=&-\frac{jk_{t}^2}{\omega}\^z\cdot(\^n\times\ee_{t})=0.
\end{eqnarray*}
The proof of Lemma 3 is thus completed. $\qed$

Based on the conclusion of Lemma 2 and Lemma 3, we can achieve the following.

{\bf{Theorem 2:}} If the medium in the waveguide satisfies Condition $\mathfrak{C}$, then there exist independent TE modes in this waveguide, and the propagation constants $k_{z}$ obtained from employing both the longitudinal scalar magnetic field $h_{z}$ stimulation and the transverse vector electric field $\ee_{t}$ stimulation are the same.

\subsection{4.2 Independent TM Modes}

The governing equations for simulating independent TM modes by using longitudinal component $e_z$ reads:

\indent Seek $k_{z}\in{\mathbb{R}}$, $e_z\neq 0$, such that
\begin{equation} \label{3eqsm1}
\left\{ \begin{aligned}
         -\divt(\d{\ep}_{t}\nabla_{t}e_{z})&=(k^2-k_{z}^2)\ep_{zz}e_z=k_{t}^2\ep_{zz}e_z~~\mbox{in}~\Gamma, \\
         e_z&=0~~\mbox{on}~\partial\Gamma.
        \end{aligned} \right.
\end{equation}
 Once any eigen-pair $(k_{z},e_{z})$ in PDE (\ref{3eqsm1}) is solved, then
 \begin{eqnarray*}
\ee_{t}
=-\frac{jk_{z}}{k_{t}^2}\nabla_{t}e_{z},\quad
\h_{t}
=-\frac{j\omega}{k_{t}^2}\^z\times(\d{\ep}_{t}\nabla_{t}e_{z}).
\end{eqnarray*}

To simulate independent TM modes by using transverse component $\h_t$, the governing equations can be written as:

\indent Find $k_{z}\in{\mathbb{R}}$, $\h_t\neq 0$, such that
\begin{equation} \label{3eqsm2}
\left\{ \begin{aligned}
         \curlt\big(\ep_{zz}^{-1}\curlt\h_t\big)-\omega^2\d{\mu}_{t}\h_{t}&
         =k_{z}^2\big(\^z\times\d{\ep}_{t}^{-1}\^z\times\h_t\big)~~\mbox{in}~\Gamma, \\
         \divt(\d{\mu}_{t}\h_t)&=0~~\mbox{in}~\Gamma,\\
         \^n\times\big(\ep_{zz}^{-1}\curlt\h_t\big)&=0~~\mbox{on}~\partial\Gamma,\\
         \^n\cdot(\d{\mu}_{t}\h_t)&=0~~\mbox{on}~\partial\Gamma.
        \end{aligned} \right.
\end{equation}
 Once any eigen-pair $(k_{z},\h_{t})$ in PDE (\ref{3eqsm2}) is solved, then
 \begin{eqnarray}
  \ee_{t}=-\frac{k_{z}\d{\ep}_{t}^{-1}(\^z\times\h_t)}{\omega},\label{boundary44}\\
e_z=\^z\cdot\frac{\ep_{zz}^{-1}\curlt\h_t}{j\omega}.\label{boundary45}
\end{eqnarray}

Similarly, we can obtain the following Lemma 4 and Lemma 5 from equations (\ref{3eqsm1}) and (\ref{3eqsm2}).

\noindent{\bf{Lemma 4.}} If $(k_{z},e_{z})$ is an eigen-pair of PDE (\ref{3eqsm1}),  then $(k_{z}, -\frac{j\omega}{k_{t}^2}\^z\times(\d{\ep}_{t}\nabla_{t}e_{z}))$ is also an eigen-pair of PDEs (\ref{3eqsm2}). \\
Proof: About the verification of the first two equations of PDEs (\ref{3eqsm2}), this step is the same as the case in the independent TE modes, therefore we omit this step. Next we examine the two boundary conditions in PDEs (\ref{3eqsm2})
:
\begin{eqnarray*}
  &&\^n\times\big(\ep_{zz}^{-1}\curlt\h_t\big) = \^n\times\bigg(\ep_{zz}^{-1}\curlt\big(-\frac{j\omega}{k_{t}^2}\^z\times(\d{\ep}_{t}\nabla_{t}e_{z})\big)\bigg) \\
   &=&\frac{j\omega\ep_{zz}^{-1}}{k_{t}^2}\^n\times\big(-\^z\divt(\d{\ep}_{t}\gradt e_{z}) \big)=\frac{j\omega\ep_{zz}^{-1}}{k_{t}^2}\^n\times\^z(k_{t}^2\ep_{zz} e_{z})\\
   &=&j\omega\^n\times\^z e_{z}=0\mbox{~on~}\partial\Gamma.
\end{eqnarray*}
Because $\^n\times\E=0,~e_{z}=0\mbox{~on~}\partial\Gamma$, then $\^n\times\ee_{t}=0\mbox{~on~}\partial\Gamma$ holds:
\begin{eqnarray*}
 &~&\^n\cdot(\d{\mu}_{t}\h_t)=\^n\cdot\bigg(\d{\mu}_{t}\big(-\frac{j\omega}{k_{t}^2}\^z\times(\d{\ep}_{t}\nabla_{t}e_{z})\big)\bigg)\\
 &=&-\frac{j\omega}{k_{t}^2}\^n\cdot\big(\d{\mu}_{t}\^z\times(\d{\ep}_{t}\gradt e_{z})\big)=\frac{j\omega}{k_{t}^2}\^n\cdot\big(\^z\times\^z\times\d{\mu}_{t}\^z\times(\d{\ep}_{t}\gradt e_{z})\big)\\
 &=&\frac{j\omega}{k_{t}^2}\^n\cdot\big(\^z\times(\^z\times\d{\mu}_{t}\^z\times\d{\ep}_{t})\gradt e_{z}\big)=\frac{j\omega}{k_{t}^2}\^n\cdot\big(\^z\times(-\frac{k^2}{\omega^2})\gradt e_{z}\big)\\
 &=&\frac{jk^2}{\omega k_{t}^2}\^z\cdot(\^n\times\gradt e_{z})=-\frac{k^2}{\omega k_{z}}\^z\cdot(\^n\times\ee_{t})=0\mbox{~on~}\partial\Gamma.
\end{eqnarray*}
This completes the proof of Lemma 4. $\qed$

\noindent{\bf{Lemma 5.}} If $(k_{z},\h_{t})$ is an eigen-pair of PDEs (\ref{3eqsm2}), then $(k_{z}, \^z\cdot\frac{\ep_{zz}^{-1}\curlt\h_t}{j\omega})$ is also an eigen-pair of PDE (\ref{3eqsm1}).\\
Proof: About the verification of the equation in PDE (\ref{3eqsm1}), this step is the same as the case in the independent TE modes, therefore we omit this step. Next we examine the boundary conditions in PDE (\ref{3eqsm1}). According to the equation (\ref{boundary44}), we have
\begin{eqnarray*}
&~&\^n\times\ee_{t}=
\^n\times\bigg(-\frac{k_{z}}{\omega}\d{\ep}_{t}^{-1}\^z\times\h_{t}\bigg)
=\frac{k_{z}}{\omega}\^n\times\bigg(\^z\times\^z\times\d{\ep}_{t}^{-1}\^z\times\h_{t}\bigg)\\
&=&\frac{k_{z}}{\omega}\^n\times\bigg(\^z\times\d{\mu}_{t}\big(\d{\mu}_{t}^{-1}\^z\times\d{\ep}_{t}^{-1}\^z\times\big)\h_{t}\bigg)
=-\frac{\omega k_{z}}{k^2}\^n\times\big(\^z\times\d{\mu}_{t}\h_{t}\big)\\
&=&-\frac{\omega k_{z}}{k^2}\bigg(\^z(\^n\cdot(\d{\mu}_{t}\h_{t}))-\d{\mu}_{t}\h_{t}(\^n\cdot\^z)\bigg)=0,
\end{eqnarray*}
By virtue of $\^n\times\E=0$ on $\partial\Gamma$, then we have $\^n\times(\ee_{t}+\^ze_{z})e^{-jk_{z}z}=0$ on $\partial\Gamma$. From this equation, then we obtain $\^n\times(\^ze_{z})=0$ on $\partial\Gamma$, thus $e_z=0$ on $\partial\Gamma$. In addition, according to (\ref{boundary45}), we have
\begin{eqnarray*}
 \^n\times\^ze_z =\frac{1}{j\omega}\^n\times(\ep_{zz}^{-1}\curlt\h_{t})=0\mbox{~on~}\partial\Gamma,
\end{eqnarray*}
therefore we prove $e_z=0\mbox{~on~}\partial\Gamma$ again, which completes the proof of Lemma 5. $\qed$

Based on the conclusion of Lemma 4 and Lemma 5, we can achieve the following\\
{\bf{Theorem 3:}} If the medium in the waveguide satisfies Condition $\mathfrak{C}$, then there exist independent TM modes in this waveguide, and the propagation constants $k_{z}$ obtained from employing both the longitudinal scalar electric field $e_{z}$ stimulation and the transverse vector magnetic field $\h_{t}$ stimulation are the same.

\section{5. Conclusion}

According to the above discussions, we can achieve the following important\\
{\bf{Theorem 4:}} There exist independent TE modes and independent TM modes in the PEC waveguide filled with a homogeneous lossless anisotropic medium if and only if the medium parameters in this waveguide satisfies Condition $\mathfrak{C}$:
\begin{eqnarray}\label{eq:34}
\left\{
  \begin{array}{ll}
  \d{\ep}=
   \begin{bmatrix} \ep & (b+cj)\ep & 0\\ (b-cj)\ep &  a^2\ep & 0 \\ 0 & 0& \ep_{zz}\end{bmatrix}
   ,~\d{\mu}=\begin{bmatrix} \mu & (b-cj)\mu & 0 \\(b+cj)\mu & a^2\mu & 0\\ 0 & 0& \mu_{zz} \end{bmatrix}, \\
   \ep>0, \mu>0, \ep_{zz}>0, \mu_{zz}>0, a^{2}-(b^{2}+c^{2})>0.
  \end{array}
\right.
\end{eqnarray}

{\bf{Remarks:}} When the waveguide is filled with a homogenous lossless isotropic medium,
then there are independent TE and TM modes in this waveguide.
In this case, $a^2=1$, $b=0$, $c=0$, $\ep=\ep_{zz}>0$ and $\mu=\mu_{zz}>0$.
Obviously, the medium parameters have already satisfied the above Condition $\mathfrak{C}$.

Theorem 4 implies that the waveguide has independent TE modes if and only if the waveguide has independent TM modes.
The case where the waveguide has independent TE modes but does not have independent TM modes does not exist. Similarly, the case where the waveguide has independent TM modes but does not have independent TE modes also dose not exist.

Future work will demonstrate the applications of this necessary and sufficient condition.

\begin{acknowledgments}
{\bf Acknowledgments}\\
This work was supported by the National Science Foundation of China under Grant 11361013 (Tian Xia), Grant 11101381, and Grant 41390453, and by the Fujian Province Natural Science
Foundation under Grant 2013J05060.
\end{acknowledgments}

\end{document}